\documentclass[12pt]{article}
\usepackage{amssymb,amsmath,epsfig}

\begin{document}

\title{\bf Mechanical Stability of Cylindrical Thin-Shell Wormholes}

\author{M. Sharif$^1$ \thanks{msharif.math@pu.edu.pk} and M. Azam$^{1,2}$
\thanks{azammath@gmail.com}\\
$^1$ Department of Mathematics, University of the Punjab,\\
Quaid-e-Azam Campus, Lahore-54590, Pakistan.\\
$^2$ Division of Science and Technology, University of Education,\\
Township Campus, Lahore-54590, Pakistan.}

\date{}

\maketitle
\begin{abstract}
In this paper, we apply the cut and paste procedure to charged black
string for the construction of thin-shell wormhole. We consider the
Darmois-Israel formalism to determine the surface stresses of the
shell. We take Chaplygin gas to deal with the matter distribution on
shell. The radial perturbation approach (preserving the symmetry) is
used to investigate the stability of static solutions. We conclude
that stable static solutions exist both for uncharged and charged
black string thin-shell wormholes for particular values of the
parameters.
\end{abstract}
{\bf Keywords:} Israel junction conditions; Stability; Black strings.\\
{\bf PACS:} 04.20.Gz; 04.40.Nr; 98.80.Jk.

\section{Introduction}

The study of thin-shell wormholes is of great interest as it links
two same or different universes by a tunnel (throat) \cite{1}. The
throat is threaded by the unavoidable amount of exotic matter (which
violates the null energy condition). It is necessary to minimize the
violation of energy conditions for the physically viability of
wormholes. For this purpose, cut and paste procedure was considered
by various authors, for instance \cite{2}-\cite{4}, to construct a
theoretical wormhole (thin-shell). Moreover, the required exotic
matter to support wormhole can be minimized with the appropriate
choice of geometry \cite{5}.

The key issue of thin-shell wormhole is to explored its mechanical
stability under radial perturbations in order to understand its
dynamical aspects. In this scenario, many people worked for the
stability analysis of thin-shell wormholes. Poisson and Visser
\cite{5a} have constructed the Schwarzschild thin-shell wormhole and
explored its stability regions under linear perturbations. Visser
\cite{6} analyzed stability of thin-shell wormholes with specific
equations of state. It was found that charge \cite{10} and positive
cosmological constant \cite{11} increase the stability regions of
spherically symmetric thin-shell wormholes. Thibeault et al.
\cite{12} explored stability of thin-shell wormhole in
Einstein-Maxwell theory with a Gauss-Bonnet term.

Cylindrical thin-shell wormholes associated with and without cosmic
strings have been widely discussed in literature
\cite{13}-\cite{16}. These cosmic strings have many astrophysical
phenomena like structure formation in the early universe and
gravitational lensing effects \cite{17}. Eiroa and Simeone \cite{18}
have discussed the cylindrical thin-shell wormholes associated with
local and global cosmic strings and found that the wormhole
configurations are unstable under velocity perturbations. Bejarano
et al. \cite{19} discussed stability of static configurations of
cylindrical thin-shell wormholes under perturbations and found that
the throat will expand or collapse depending upon the sign of the
velocity perturbations.

Richarte and Simeone \cite{20} found more configurations which are
not stable under radial velocity perturbations, consistent with the
conjecture discussed in a paper \cite{19}. Eiroa and Simeone
\cite{21} summarized the above results and found that stable
configurations are not possible for the cylindrical thin-shell
wormholes. Recently, we have explored stability of spherical and
cylindrical geometries at Newtonian and post-Newtonian
approximations and also spherically symmetric thin-shell wormholes
\cite{21a}.

Several candidates have been proposed like quintessence, K-essence,
phantom, quintom, tachyon, family of Chaplygin gas, holographic and
new agegraphic DE \cite{29} for the explanation of accelerated
universe. Besides all these candidates, Chaplygin gas (an exotic
matter) has been proposed widely to explain the expansion of the
universe. Kamenshchik et al. \cite{30} described the feature of
Chaplygin gas and explored cosmology of FRW universe filled with a
Chaplygin gas. The Chaplygin cosmological models support the
observational evidence \cite{31,32}.

Models of exotic matter like phantom energy with equation of state
$(p=\omega\sigma,~ \omega<-1)$ \cite{35} and Chaplygin gas
$(p\sigma=-A)$ \cite{36}, where $A$ is a positive constant, has been
of interest in wormhole construction. Wormholes have also been
studied in dilaton gravity \cite{37} and Einstein-Gauss-Bonnet
theory \cite{38}. Eiroa \cite{39} found stable static solutions of
spherically symmetric thin-shell wormholes with Chaplygin equation
of state.

In this paper, we construct black string thin-shell wormholes with
and without charge supported by the Chaplygin gas. We study the
mechanical stability of these constructed wormholes under radial
perturbations preserving the cylindrical symmetry. The format of the
paper is as follows. In section \textbf{2}, we formulate surface
stresses of the matter localized on the shell through Darmois-Israel
junction conditions. Section \textbf{3} provides the mechanical
stability analysis of the static configuration of thin-shell
wormholes. In the last section \textbf{4}, we conclude our results.

\section{Thin-Shell Wormhole Construction}

In this section, we build a thin-shell wormhole from charged black
string through cut and paste technique and discuss its dynamics
through the Darmois-Israel formalism. We consider the
Einstein-Hilbert action with electromagnetic field as \cite{a}
\begin{eqnarray}\label{a}
\mathcal{A}+\mathcal{A}_{em}=\frac{1}{16\pi{G}}\int{\sqrt{-g}(R-2\Lambda)d^4x}
-\frac{1}{16\pi}\int{\sqrt{-g}F^{\mu\nu}F_{\mu\nu}d^4x},
\end{eqnarray}
where $\mathcal{A}$ is defined by
\begin{eqnarray}\label{b}
\mathcal{A}=\frac{1}{16\pi{G}}\int{\sqrt{-g}(R-2\Lambda)d^4x}.
\end{eqnarray}
Here, $g,~R,~\Lambda,~F_{\mu\nu}$ are the metric determinant, the
Ricci scalar, negative cosmological constant and the Maxwell field
tensor ($F_{\mu\nu}=\partial_\mu{\phi_\nu}-\partial_\nu{\phi_\mu}$),
respectively and $\phi_\nu=-\lambda(r)\delta^0_\nu$ is the
electromagnetic four potential with an arbitrary function
$\lambda(r)$.

The Einstein-Maxwell equations from the above action yields a
cylindrically symmetric vacuum solution, i.e., charged black string
given as
\begin{equation}\label{1}
ds^2=-g(r)dt^{2}+g^{-1}(r)dr^{2}+h(r)(d\phi^{2}+\alpha^2{dz^2}),
\end{equation}
where
$g(r)=\left(\alpha^2r^2-\frac{4M}{\alpha{r}}+\frac{4Q^2}{\alpha^2r^2}\right)$
is a positive function for the given radius and $h(r)=r^2$. We have
following restrictions on coordinates to preserve the cylindrical
geometry
$$-\infty<t<\infty,\quad 0\leq{r}<\infty, \quad
-\infty<{z}<{\infty},\quad 0\leq{\phi}\leq{2\pi}.$$ The parameters
$M,~Q$ are the ADM mass and charge density, respectively and
$\alpha^2=-\frac{\Lambda}{3}>0$. We remove the region with $r<a$ of
the given cylindrical black string and take two identical $4D$
geometries $V^{\pm}$ with $r\geq{a}$ as
\begin{equation}\label{2}
V^{\pm}=\{x^{\gamma}=(t,r,\phi,z)/r\geq{a}\},
\end{equation}
where $``a"$ is the throat radius and glue these geometries at the
timelike hypersurface $\Sigma=\Sigma^\pm=\{r-a=0\}$ to get a new
manifold $V=V^{+}\cup{V^{-}}$. This manifold is geodesically
complete exhibiting a wormhole with two regions connected by a
throat satisfying the radial flare-out condition, i.e., $h'(a)=2a>0$
\cite{16}. If $`{r_h}$' is the event horizon of the black string
given in Eq.(\ref{1}), then we assume $a>r_h$ in order to prevent
the occurrence of horizons and singularities in wormhole
configuration.

We use the standard Darmois-Israel formalism \cite{b,c} to analyze
the dynamics of the wormhole. The two sides of the shell are matched
through the extrinsic curvature defined on $\Sigma$ as
\begin{equation}\label{3}
K^{\pm}_{ij}=-n^{\pm}_{\gamma}(\frac{{\partial}^2x^{\gamma}_{\pm}}
{{\partial}{\xi}^i{\partial}{\xi}^j}+{\Gamma}^{\gamma}_{{\mu}{\nu}}
\frac{{{\partial}x^{\mu}_{\pm}}{{\partial}x^{\nu}_{\pm}}}
{{\partial}{\xi}^i{\partial}{\xi}^j}),\quad(i, j=0,2,3),
\end{equation}
where $\xi^i=(\tau,\theta,\phi)$ are the coordinates on $\Sigma$ and
$n^{\pm}_{\gamma}$ are the unit normals obtained to $\Sigma$ as
\begin{equation}\label{3a}
n^{\pm}_{\gamma}=\left(-\dot{a},\frac{\sqrt{g(r)+\dot{a}^2}}{g(r)},0,0\right),
\end{equation}
satisfying the relation $n^{\gamma}n_{\gamma}=1$. The induced metric
on $\Sigma$ is defined as
\begin{equation}\label{4}
ds^2=-d\tau^2+a^2(\tau)(d\phi^2+\alpha^2dz^2),
\end{equation}
where $\tau$ is a the proper time on the hypersurface. The
non-vanishing components of the extrinsic curvature turns out to be
\begin{equation}\label{5}
K^{\pm}_{\tau\tau}=\mp\frac{g'(a)+2\ddot{a}}{2\sqrt{g(a)+\dot{a}^2}},
\quad K^{\pm}_{\phi\phi}= \pm{a}\sqrt{g(a)+\dot{a}^2},\quad
K^{\pm}_{zz}=\alpha^2K^{\pm}_{\phi\phi}.
\end{equation}
Here dot and prime mean derivative with respect to $\tau$ and $r$
respectively.

The surface stress-energy tensor $S_{ij}=diag(-\sigma,~p,~p)$
provides surface energy density $\sigma$ and surface tensions $p$ of
the shell. The Lanczos equations are defined on the shell as
\begin{equation}\label{6}
S^i_{j}=-\frac{1}{8\pi}\left[\kappa^i_{j}-{\delta}^i_j{\kappa}^k_k\right],
\end{equation}
where $\kappa_{ij}=K^{+}_{ij}-K^{-}_{ij}$ and due to the simplicity
of cylindrical symmetry, we have
$\kappa^i_j=diag(\kappa^\tau_{\tau},~\kappa^\phi_{\phi},~\kappa^\phi_{\phi})$.
The Lanczos equations with surface stress-energy tensor provides
\begin{eqnarray}\label{7}
\sigma&=&-\frac{1}{4\pi}\kappa^\phi_\phi=-\frac{1}{2\pi{a}}\sqrt{g(a)+\dot{a}^2},\\\label{8}
p&=&\frac{1}{8\pi}(\kappa^\tau_\tau+\kappa^\phi_\phi)=\frac{1}{8\pi{a}}\frac{2a\ddot{a}+2\dot{a}^2
+2g(a)+ag'(a)}{\sqrt{g(a)+\dot{a}^2}}.
\end{eqnarray}
\textit{Matter that violates the null energy condition is known as
exotic matter}. We see from Eq.(\ref{7}) that the surface energy
density is negative indicating the existence of exotic matter at the
throat. For the explanation of such matter, the Chaplygin equation
of state is defined on the shell as
\begin{equation}\label{9}
p=-\frac{A}{\sigma},
\end{equation}
where $A>0$. Using Eqs.(\ref{7}) and (\ref{8}) in the above
equation, we obtain a second order differential equation satisfied
by the throat radius $`a$'
\begin{equation}\label{10}
2a\ddot{a}+2\dot{a}^2-16\pi^2{A}a^2+2g(a)+ag'(a)=0.
\end{equation}

\section{Stability Analysis}

In this section, we have adapted and applied the criteria introduced
in Ref.\cite{39} for the stability analysis of static solutions. The
existence of static solutions is subject to the condition $a_0>r_h$.
For such solutions, we consider static configuration of
Eq.(\ref{10}) as
\begin{equation}\label{11}
-16\pi^2{A}a^2_0+2g(a_0)+a_0g'(a_0)=0,
\end{equation}
and the corresponding static configuration of surface energy density
and pressure are
\begin{eqnarray}\label{12}
\sigma_0=-\frac{1}{2\pi{a_0}}\sqrt{g(a_0)},\quad
p_0=\frac{2A\pi{a_0}}{\sqrt{g(a_0)}}.
\end{eqnarray}
The perturbed form of the throat radius is given by
\begin{equation}\label{14}
a(\tau)=a_0[1+\epsilon(\tau)],
\end{equation}
where $\epsilon(\tau)\ll1$ is a small perturbation preserving the
symmetry. Using the above equation, the corresponding perturbed
configuration of Eq.(\ref{10}) can be written as
\begin{equation}\label{15}
(1+\epsilon)\ddot{\epsilon}+\dot{\epsilon}^2-8\pi^2{A}(2+\epsilon)\epsilon
+D(a_0,\epsilon)=0,
\end{equation}
where
\begin{equation*}\label{16}
D(a_0,\epsilon)=\frac{2g(a_0+a_0\epsilon)+a_0(1+\epsilon)g'(a_0+a_0\epsilon)
-2g(a_0)-a_0g'(a_0)}{2a^2_0}.
\end{equation*}
Substituting $\upsilon(\tau)=\dot{\epsilon}(\tau)$ in Eq.(\ref{15}),
we have a first order differential equation in $\upsilon$
\begin{equation}\label{17}
\dot{\upsilon}=\frac{8\pi^2{A}(2+\epsilon)\epsilon-D(a_0,\epsilon)-\upsilon^2}{1+\epsilon}.
\end{equation}
Using Taylor expansion to first order in $\epsilon$ and $\upsilon$,
we obtain a set of equations
\begin{equation}\label{18}
\dot{\epsilon}=\upsilon,\quad \dot{\upsilon}=\Delta{\epsilon},
\end{equation}
where
\begin{eqnarray}\label{19}
\Delta&=&16\pi^2{A}-\frac{3g'(a_0)+a_0g''(a_0)}{2a_0}.
\end{eqnarray}
This set of equations can be written in matrix form as
$$\dot{\eta}=L\eta,$$ where $\eta$ and $L$ are
\begin{small}
\begin{equation*}
\eta=\left[\begin{array}{ccccc} \epsilon \\ \upsilon
\end{array} \right],\quad
L=\left[\begin{array}{ccccc}
0&1 \\
\Delta&0
\end{array} \right].\\
\end{equation*}
\end{small}
The matrix $`L$' has two eigenvalues $\pm{\sqrt{\Delta}}$.

The stability analysis of static solutions on the basis of
$\Delta$ can be formulated as follows \cite{39}: \\\\
\textbf{i.} When $\Delta>0$, the given matrix has two real
eigenvalues $\lambda_1=-\sqrt{\Delta}<0$ and
$\lambda_2=\sqrt{\Delta}>0$. The negative eigenvalue has no physical
significance, while the positive eigenvalue indicates the
presence of unstable static solution.\\\\
\textbf{ii.} When $\Delta=0$, the corresponding eigenvalues are
$\lambda_1=\lambda_2=0$ and Eq.(\ref{18}) leads to
$\upsilon=\upsilon_0=constant$ and
$\epsilon=\epsilon_0+\upsilon_0(\tau-\tau_0)$.
In this case, the static solution is unstable.\\\\
\textbf{iii.} When $\Delta<0$, we have two imaginary eigenvalues
$\lambda_1=-\dot{\iota}\sqrt{\mid\Delta\mid}$ and
$\lambda_2=\dot{\iota}\sqrt{\mid\Delta\mid}$. Eiroa \cite{39}
discussed this case in detail and found that the only stable static
solution with throat radius $a_0$ exists for $\Delta<0$ which is not
asymptotically stable.

\subsection{Uncharged Black String Thin-Shell Wormhole}

Here, we investigate the stability of thin-shell wormhole static
solution constructed from the asymptotically anti-de Sitter
uncharged black string \cite{a}. For the uncharged black string, we
have
\begin{equation}\label{21}
g(r)=\alpha^2r^2-\frac{4M}{\alpha{r}}.
\end{equation}
The event horizon of the black string is given by $(g_{tt}=0)$
\begin{equation}\label{21a}
r_{h}=\frac{(4M)^{\frac{1}{3}}}{\alpha}.
\end{equation}
The corresponding surface energy density and pressure of black
string for static configuration leads to
\begin{eqnarray}\label{22}
\sigma_0=-\frac{\sqrt{\alpha^3a^3_0-4M}}{2\pi{a^{\frac{3}{2}}_0}\sqrt{\alpha}},\quad
p_0=\frac{2A\pi{a^{\frac{3}{2}}_0}\sqrt{\alpha}}{\sqrt{\alpha^3a^3_0-4M}}.
\end{eqnarray}
Using Eq.(\ref{21}) in (\ref{11}), we get a cubic equation satisfied
by the throat radius with $A>0$ and $M>0$
\begin{equation}\label{23}
(4A\pi^2{\alpha}-\alpha^3)a^3_0+M=0.
\end{equation}
The corresponding roots $a'^1_0,~a'^2_0,~a'^3_0,$ of the cubic
equation are
\begin{eqnarray}\label{24}
a'^1_0&=&\frac{1}{\alpha}\left(\frac{M}{
1-4A{\pi^2}\alpha^{-2}}\right)^\frac{1}{3},\\\label{25}
a'^2_0&=&\frac{(-1+\dot{\iota}\sqrt{3})}{2\alpha}\left(\frac{M}
{1-4A{\pi^2}\alpha^{-2}}\right)^\frac{1}{3},\\\label{26}
a'^3_0&=&\frac{(-1-\dot{\iota}\sqrt{3})}{2\alpha}\left(\frac{M}
{1-4A{\pi^2}\alpha^{-2}}\right)^\frac{1}{3}.
\end{eqnarray}

Now, we analyze all the roots numerically and find which of them
represent static solution. The condition for the existence of static
solution is subject to $a'_0>\frac{(4M)^{\frac{1}{3}}}{\alpha}.$ For
$(A\alpha^{-2})>(4\pi^2)^{-1},$ we have one negative real root
$a'^3_0$ and two non-real roots $a'^1_0,~a'^2_0$. All the three
roots have no physical significance. Thus, there are no static
solutions corresponding to $(A\alpha^{-2})>(4\pi^2)^{-1}$. For
$(A\alpha^{-2})<(4\pi^2)^{-1},$ there is one positive real root
$a'^1_0$ and two non-real roots $a'^2_0,~a'^3_0$. The non-real roots
are discarded being of no physical significance. We check
numerically that for $(A\alpha^{-2})<(4\pi^2)^{-1},$ the only static
solution corresponding to positive real root is
$a'^1_0\geq{\frac{(4M)^{\frac{1}{3}}}{\alpha}}$. The critical value
for which the throat radius greater than event horizon is
$\beta=1.9\times10^{-2}<(4\pi^2)^{-1}$. It turns out that the static
solution will exist if $\beta<(A\alpha^{-2})\leq(4\pi^2)^{-1}$.

For the stability of $a'^1_0$, we work out $\Delta$ for the
uncharged black string. From Eq.(\ref{19}), we have
\begin{equation}\label{27}
\Delta=16A\pi^2-4\alpha^2-\frac{2M}{\alpha{a^3_0}}.
\end{equation}
\begin{figure}
\centering \epsfig{file=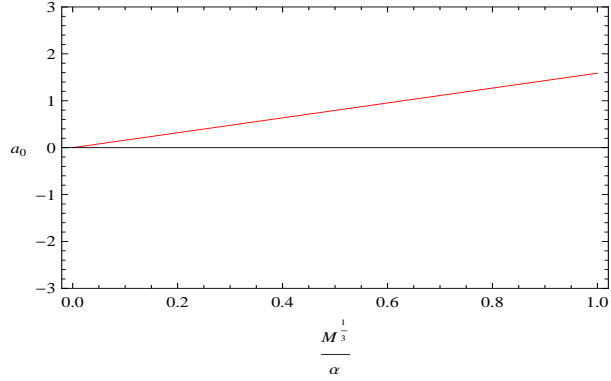,width=.58\linewidth, height=2in}
\caption{Plot of stable static solution of $a'^1_0$ for
$\beta<(A\alpha^{-2})\leq(4\pi^2)^{-1}$.}
\end{figure}
The above equation with Eq.(\ref{23}) leads to
\begin{equation}\label{28}
\Delta=-\frac{6M}{\alpha{a^3_0}}.
\end{equation}
This shows that $\Delta$ is always negative for
$a'^1_0>\frac{(4M)^{\frac{1}{3}}}{\alpha}$. Thus, we have one stable
solution for the uncharged black string corresponding to
$(A\alpha^{-2})<(4\pi^2)^{-1}$ with throat radius $a'^1_0$ as shown
in Figure \textbf{1}. There does not exist any other static solution
corresponding to $(A\alpha^{-2})>(4\pi^2)^{-1}$.

\subsection{Charged Black String Thin-Shell Wormhole}

For the case of charged black string, we have considered $g(r)$ from
Eq.(\ref{1})
\begin{equation}\label{29}
g(r)=\alpha^2r^2-\frac{4M}{\alpha{r}}+\frac{4Q^2}{\alpha^2r^2}.
\end{equation}
The event horizons \cite{a} for the charged black string are
obtained from the equation
\begin{equation}\label{29a}
\alpha^2r^2-\frac{4M}{\alpha{r}}+\frac{4Q^2}{\alpha^2r^2}=0.
\end{equation}
The above quartic equation has two real and two complex roots. We
discard the complex roots being of unphysical and the real roots are
taken as inner and outer event horizons of the charged black string
\begin{equation}\label{30}
r_{\pm}=\frac{(4M)^\frac{1}{3}}{2\alpha}\left[\sqrt{s}\pm
\sqrt{{2}\sqrt{s^2-Q^2\left(\frac{2}{M}\right)^\frac{4}{3}}-s}\right]
=\frac{(4M)^\frac{1}{3}}{\alpha}\chi,
\end{equation}
provided that the inequality $Q^2\leq\frac{3}{4}M^\frac{4}{3}$
holds, where $s$ and $\chi$ are given by
\begin{eqnarray}\label{31}
s&=&\left(\frac{1}{2}+\frac{1}{2}\sqrt{1-\frac{64Q^6}{27M^4}}
\right)^\frac{1}{3}
+\left(\frac{1}{2}-\frac{1}{2}\sqrt{1-\frac{64Q^6}{27M^4}}
\right)^\frac{1}{3},\\\label{31a}
\chi&=&\frac{1}{2}\left[\sqrt{s}\pm
\sqrt{{2}\sqrt{s^2-Q^2\left(\frac{2}{M}\right)^\frac{4}{3}}-s}\right].
\end{eqnarray}
For $Q^2>\frac{3}{4}M^\frac{4}{3}$, the given metric has no event
horizon and represents a naked singularity. If
$Q^2=\frac{3}{4}M^\frac{4}{3}$, the inner and outer horizons
coincide, which corresponds to extremal black strings. For the
static wormhole associated to the charged black string, the surface
energy density and pressure turn out to be
\begin{eqnarray}\label{32}
{\sigma_0}&=&-\frac{\sqrt{\alpha^4a^4_0-4M\alpha{a_0+4Q^2}}}{2\pi{a^2}_0{\alpha}},\\\label{32a}
p_0&=&\frac{2A\pi{a^2_0}{\alpha}}{\sqrt{\alpha^4a^4_0-4M\alpha{a_0+4Q^2}}}.
\end{eqnarray}
\begin{figure}
\centering \epsfig{file=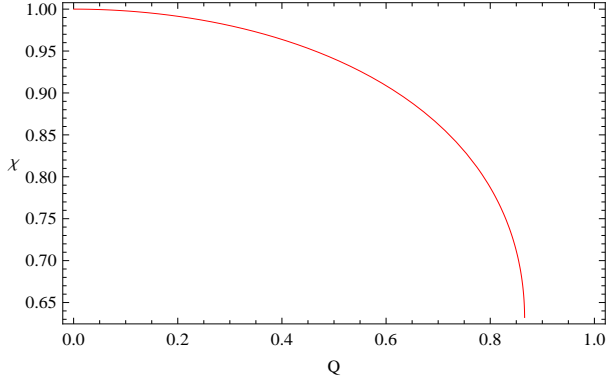,width=.58\linewidth, height=2in}
\caption{Plot of $\chi$ vs $Q$ shows that $\chi$ is a decreasing
function of $Q$.}
\end{figure}

We would like to find an equation satisfied by the throat radius. We
substitute Eq.(\ref{29}) in (\ref{11}), it turns out that the charge
terms cancel out and again we get the same equation (\ref{23})
satisfied by the throat radius of the wormhole with the roots given
in Eqs.(\ref{24})-(\ref{26}). For the existence of static solutions,
the throat radius $a_0$ should be greater than $r_h=r_+$. It is
noted that the event horizon (\ref{30}) is a decreasing function of
$Q$ for positive values of $M$ and $\alpha$ as shown in Figure
\textbf{2}. Again, we check all the roots numerically and find that
there exists only one static solution, $a'^1_0$, subject to large
values of $Q$. Thus in the case of charged black string, we have one
static solution for $(A\alpha^{-2})<(4\pi^2)^{-1}$ with $Q<M$ and
large values of $Q$ and and no static solution for
$(A\alpha^{-2})>(4\pi^2)^{-1}$. The stability of static solutions,
whether they are stable or unstable, depends upon the sign of
$\Delta$ which is again negative for $a'^1_0>r_h$. Hence, the
obtained static solution is of stable type for the static wormhole
associated with the charged black string.

\section{Summary and Discussion}

In this paper, we have constructed black string thin-shell wormholes
and investigated their mechanical stability under radial
perturbations preserving the cylindrical symmetry. We have
formulated the Darmois Israel junction conditions and imposed
Chaplygin equation of state on the matter shell. We have constructed
wormholes using cut and paste procedure such that the throat radius
should be greater than the event horizon of the given metric, i.e.,
$a_0>r_h$. A dynamical equation with the Chaplygin gas of the shell
has been obtained by considering throat radius as a function of
proper time. The proposed criteria for the stability analysis of
static solutions has been extended to uncharged and charged
constructed wormholes.

It was shown in Refs.\cite{18}-\cite{21} that cylindrical thin-shell
wormholes with equations of state depending upon the metric
functions are always mechanically unstable. On the other hand, we
have constructed cylindrical thin-shell wormholes from charged black
string solution (whose causal structure is similar to
Reissner-Nordstr$\ddot{o}$m solution) with Chaplygin equation of
state. We have found stable configurations depending upon the
parameters involving in the model. This shows that the choice of
bulk solution as well as equation of state plays a significant role
in the stability of wormhole configurations.

For the case of static wormhole associated with uncharged black
string, we have found one negative real root and two non-real roots
corresponding to $(A\alpha^{-2})>(4\pi^2)^{-1}$, which implies no
static solution. For $(A\alpha^{-2})<(4\pi^2)^{-1}$, we have two
complex roots and one positive real root which is stable for
$\beta<(A\alpha^{-2})\leq(4\pi^2)^{-1}$, where,
$\beta=1.9\times10^{-2}<(4\pi^2)^{-1}$. In the case of static
wormhole associated to charged black string, we see that event
horizon $r_h$ is a decreasing function of $Q$. Also, we have found
the same three roots as in the case of uncharged black string. Thus,
there is one stable static solution corresponding  to
$(A\alpha^{-2})<(4\pi^2)^{-1}$ depending upon the values of the
parameters $Q<M$ and $\alpha>0$. For $(A\alpha^{-2})>(4\pi^2)^{-1}$,
there is no static solution. It is mentioned here that there always
exists one stable static solution for $(A\alpha^{-2})<(4\pi^2)^{-1}$
and no static solution for $(A\alpha^{-2})>(4\pi^2)^{-1}$ in both
uncharged as well as charged black string.

\vspace{0.5cm}

{\bf Acknowledgments}

\vspace{0.5cm}

We would like to thank the Higher Education Commission, Islamabad,
Pakistan, for its financial support through the {\it Indigenous
Ph.D. 5000 Fellowship Program Batch-VII}. One of us (MA) would like
to thank University of Education, Lahore for the study leave.

\end{document}